\documentclass[9pt,twocolumn,twoside]{IEEEtran}

\usepackage[cmex10]{amsmath}
\usepackage{cite}

\usepackage{latexsym}
\usepackage{graphics}
\usepackage{amssymb}
\usepackage{amsthm}

\usepackage{cite}
\usepackage{array}
\usepackage{mdwmath}
\usepackage{mdwtab}
\usepackage[tight,footnotesize]{subfigure}
\usepackage{graphicx}

\usepackage{stfloats}
\fnbelowfloat
\usepackage{url}

\ifCLASSOPTIONcompsoc
\usepackage[tight,normalsize,sf,SF]{subfigure}
\else
\usepackage[tight,footnotesize]{subfigure}
\fi

\def\n0{n_{0}}

\def\tr{\mathrm{tr}}
\def\rank{\mathrm{rank}}

\newcommand{\fracSum}[1]{{\underset{{#1}}{\sum}}}

\newcommand{\fracSumtwo}[2]{\overset{#2}{\underset{#1}{\sum}}}

\newcommand{\vect}[1]{\mathbf{#1}}

\newcommand{\maximize}[1]{{\underset{{#1}}{\mathrm{maximize}}}}

\theoremstyle{remark}

\newtheorem{theorem}{Theorem}
\newtheorem{corollary}{Corollary}
\newtheorem{lemma}{Lemma} 
\newtheorem{definition}{Definition}

 {\begin{center}\begin{minipage}{#1}\hrule\medskip}
 {\vspace{-1ex}\hrule \end{minipage}\end{center}}

\newtheorem{strategy}{Strategy}

\begin{document}

\title{Pareto Characterization of the Multicell MIMO Performance Region With Simple Receivers}

\author{Emil Bj\"ornson,~\IEEEmembership{Member,~IEEE,}
        Mats Bengtsson,~\IEEEmembership{Senior Member,~IEEE,}
        and~Bj\"orn~Ottersten,~\IEEEmembership{Fellow,~IEEE}
\thanks{Copyright (c) 2012 IEEE. Personal use of this material is permitted. However, permission to use this material for any other purposes must be obtained from the IEEE by sending a request to pubs-permissions@ieee.org. The associate editor coordinating the review of this manuscript
and approving it for publication was Dr.~Milica Stojanovic. The research leading to these results has received funding from the European Research Council under the European Community's Seventh Framework Programme (FP7/2007-2013) / ERC grant agreement number 228044.}%
\thanks{E.~Bj\"ornson and M.~Bengtsson are with the Signal Processing Laboratory, ACCESS
Linnaeus Center, KTH Royal Institute of Technology, SE-100 44
Stockholm, Sweden (e-mail: emil.bjornson@ee.kth.se; mats.bengtsson@ee.kth.se).}%
\thanks{B.~Ottersten is with the Signal Processing Laboratory, ACCESS
Linnaeus Center, KTH Royal Institute of Technology, SE-100 44
Stockholm, Sweden. He is also with the Interdisciplinary Centre for Security, Reliability and Trust (SnT), University of Luxembourg,
6 rue Richard Coudenhove-Kalergi, L-1359 Luxembourg-Kirchberg, Luxembourg (email: bjorn.ottersten@ee.kth.se).}%
\thanks{Digital Object Identifier 10.1109/TSP.2012.2199309}
}

\markboth{IEEE TRANSACTIONS ON SIGNAL PROCESSING}%
{Bj\"ornson \MakeLowercase{\textit{et al.}}: IEEE TRANSACTIONS ON
SIGNAL PROCESSING}

\maketitle

\begin{abstract}
We study the performance region of a
general multicell downlink scenario with multiantenna transmitters, hardware impairments, and
low-complexity receivers that treat interference as noise.
The Pareto boundary of this region describes
all efficient resource allocations, but is generally hard to
compute. We propose a novel explicit characterization that gives Pareto optimal transmit strategies
using a set of positive parameters---fewer than in prior work.
We also propose an implicit characterization that requires even fewer parameters
and guarantees to find the Pareto boundary for every choice of parameters, but at the expense of solving quasi-convex optimization problems.
The merits of the two characterizations are illustrated for interference channels and ideal network multiple-input multiple-output (MIMO).
\end{abstract}


\section{Introduction}

Interference limits the performance of multicell systems.
Conventionally, interference is managed by dividing the available
frequency resources such that adjacent cells use different
subcarriers. In modern multiantenna systems, interference can
instead be managed spatially
\cite{Gesbert2010a},
potentially leading to large improvements through global reuse of
all frequency resources. However, spatial interference management
requires reliable channel state information (CSI) and joint transmission optimization across cells. This coordination is termed \emph{network MIMO}
or \emph{coordinated multi-point} (CoMP).

Multicell performance optimization is more involved than oblivious
single-cell optimization; user-fairness and inter-user interference bring a strong intercell coupling.
All achievable combinations of user throughput are ideally described by the
capacity region, derived in \cite{Weingarten2006a} when all base stations act as a single transmitter.
In practice, the capacity region is an optimistic performance measure as it, for example, relies
on global transceiver design and complex signal processing.
Herein, we consider an alternative that we call the
\emph{performance region} with simple
receivers that treat co-user interference as noise and have a single effective antenna.\footnote{It is usually called the \emph{rate
region} in prior work, but also other performance measures than the achievable data rate are considered herein.} 
These assumptions recognize the low-complexity constraints on practical user terminals.

The performance region is characterized by its \emph{Pareto boundary}, a subset of the outer
boundary where the performance cannot be improved for any user without degrading for others.
This boundary represents all efficient resource allocations.
In general, it is hard to find points on the Pareto
boundary, but there are characterizations that reduces the search-space to a few parameters.
Some characterizations are \emph{explicit}, thus they provide closed-form transmit strategies based on
the parameters \cite{Bjornson2010c, Mochaourab2011a,Jorswieck2008b,Shang2010a,Mochaourab2012a,Zhang2010a}.
Their downside is that they only provide necessary conditions; not all parameter choices achieve the Pareto boundary.
On the contrary, \emph{implicit} characterizations guarantee that a point on the outer boundary is found
\cite{Mohseni2006a,Zhang2010a,Karipidis2010a,Bjornson2012a}, but a quasi-convex problem must be solved for each choice of parameter values to find the corresponding transmit strategy.

Explicit Pareto characterizations for general multicell systems with $K_t$ transmitters and $K_r$ users were given in
\cite{Bjornson2010c} and \cite{Mochaourab2011a}. Both required $K_t(K_r-1)$ parameters for beamforming (and often
additional power-control parameters), and these were improved from complex-valued in \cite{Bjornson2010c} to $[0,1]$-parameters in \cite{Mochaourab2011a}. The special case of the multiple-input single-output (MISO)
interference channel, where each transmitter serves a single unique user, has received particular attention.
A characterization with $K_t(K_r-1)$ complex-valued parameters was derived in \cite{Jorswieck2008b} and it was improved in \cite{Shang2010a} to $[0,1]$-parameters.
A similar\footnote{Positive parameters are used in \cite{Zhang2010a}, but there are bijective functions from $[0,\infty)$ to $[0,1]$ making the complexity identical.}
characterization was proposed in \cite{Zhang2010a}, although closed-form beamforming is not guaranteed. Recently, \cite{Mochaourab2012a} showed that only a single parameter is  required for the two-user MISO interference channel (and it even ensures attaining the Pareto boundary).

Implicit Pareto characterizations have received considerably less attention. The broadcast capacity region was considered in \cite{Mohseni2006a}, and
the performance region of MISO interference channels in \cite{Zhang2010a,Karipidis2010a}. In both cases, $K_r-1$ parameters were used to define a ray from
the origin. The intersection between this ray and the outer boundary of the region was formulated as a quasi-convex optimization problem. There are connections to the
optimization of worst-user performance \cite{Wiesel2006a,Bjornson2012a}, which represents a certain ray direction.

Herein, we consider a general multicell scenario with dynamic
cooperation clusters \cite{Bjornson2011a}, basically describing anything from interference channels to ideal network MIMO in a unified manner.
The performance is measured by arbitrary functions of the single-to-noise-and-interference ratios (SINRs), we consider arbitrary linear power constraints and hardware impairments. The main contributions are as follows.

\begin{itemize}

\item A novel explicit Pareto boundary characterization is proposed. It exploits the idea of uplink-downlink duality \cite{Wiesel2006a,Yu2007a,Bjornson2011a} and uses
$K_r+L-2$ positive parameters, where $L$ is the number of power constraints. Thus, it requires fewer parameters than the prior work in \cite{Bjornson2010c,Mochaourab2011a,Jorswieck2008b,Shang2010a,Zhang2010a}.

\item The implicit Pareto boundary characterizations in \cite{Mohseni2006a,Zhang2010a,Karipidis2010a,Bjornson2012a} are extended to general multicell scenarios, using $K_r-1$ parameters in the interval $[0,1]$. Each point is given by solving a quasi-convex optimization problem.

\item The proposed characterizations include physical hardware impairments that could distort the transmitted signals, while the prior work in \cite{Gesbert2010a,Weingarten2006a,Bjornson2010c,Mochaourab2011a,Jorswieck2008b,Shang2010a,Mochaourab2012a,Mohseni2006a,Zhang2010a,Karipidis2010a,Wiesel2006a,Bjornson2012a,Bjornson2011a,Yu2007a,Annapureddy2011a,Zhang2008a} are limited to ideal transceiver hardware.
\end{itemize}

\section{System Model and Preliminaries} \label{section_systemmodel}

We consider downlink transmission with $K_t$ base
stations and $K_r$ users. The $j$th base station is denoted
$\textrm{BS}_j$ and has $N_j$ antennas. The $k$th user is denoted
$\textrm{MS}_k$ and is a simple receiver.
\begin{definition} \label{def_simple_receiver}
A \emph{simple receiver} has the following characteristics:
\begin{itemize}
\item It is viewed to have single effective antenna in the transmit optimization;
\item It treats co-user interference as noise (i.e., without trying to decode and subtract interfering signals).
\end{itemize}
\end{definition}
The first property means that $\textrm{MS}_k$ either has a single antenna or has $M_k>1$ antennas that are combined into a single effective antenna (using receive combining)
prior to transmit optimization to enable practical non-iterative transmission design. The second property means single user detection and enables low-complexity reception,
but is suboptimal except in some low-interference regimes \cite{Annapureddy2011a}.

In a general multicell scenario, some users are served in a coordinated manner by multiple transmitters. In addition, some transmitters and
receivers are very far apart, making it hard to estimate these channels and pointless to use such uncertain estimates for interference coordination.
To capture these properties and enable unified analysis, we adopt the dynamic coordination
framework of \cite{Bjornson2011a}.

\begin{definition} \label{def_dynamic_clusters}
\emph{Dynamic cooperation clusters} means that $\textrm{BS}_j$ has the following characteristics:
\begin{itemize}
\item It has perfect CSI to receivers in $\mathcal{C}_j
\subseteq \{1,\ldots,K_r\}$, while interference generated to receivers $\bar{k} \not \in \mathcal{C}_j$ are treated as additive complex Gaussian noise;
\item It serves the receivers in $\mathcal{D}_j \subseteq
\mathcal{C}_j$ with data.
\end{itemize}
\end{definition}

\begin{figure}[t!]
\begin{center}
\includegraphics[width=\columnwidth]{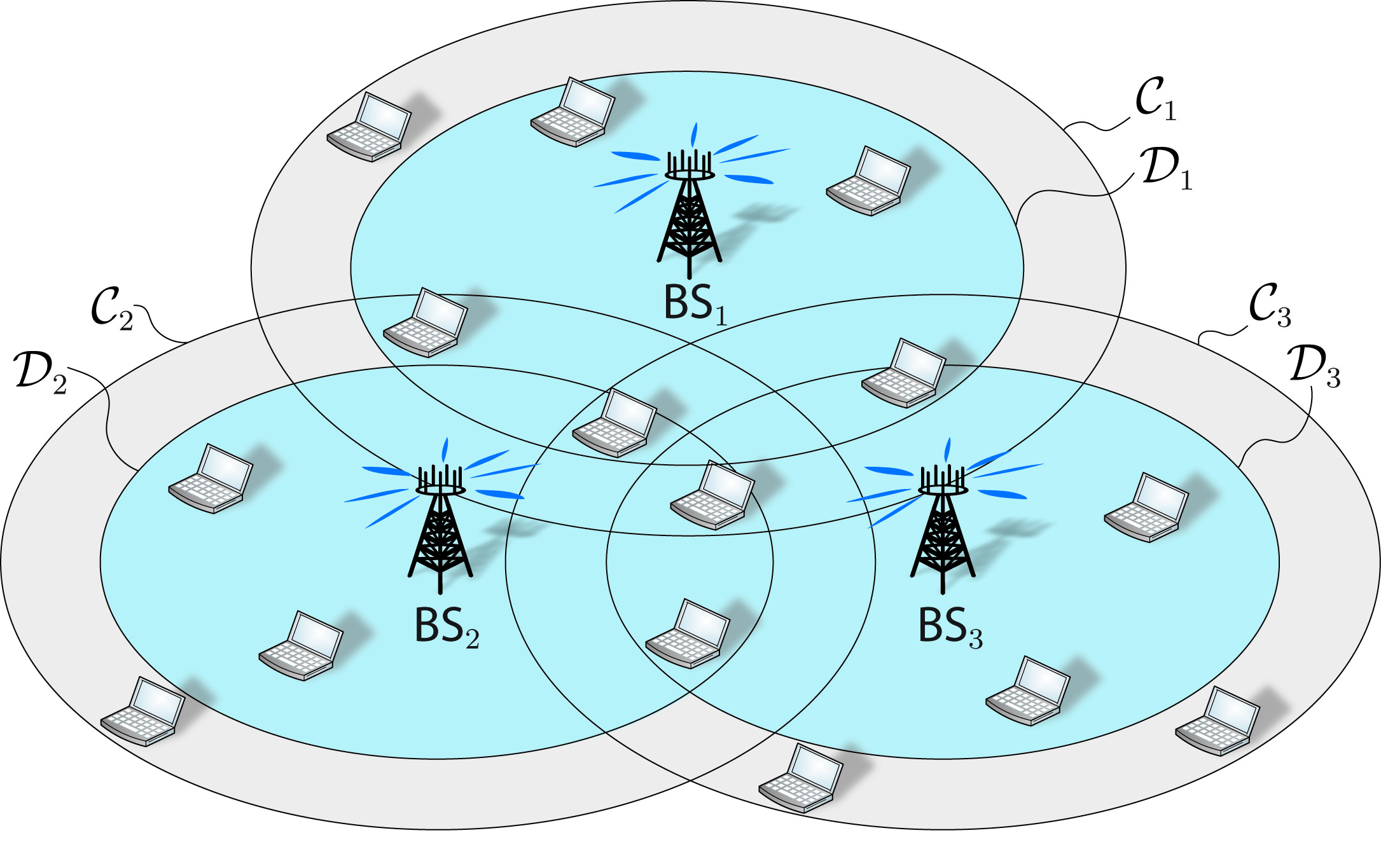}
\end{center} \vskip-3mm
\caption{Schematic intersection between three cells. $\textrm{BS}_j$
serves users in the inner circle ($\mathcal{D}_j$), while it coordinates interference to  users in the outer circle ($\mathcal{C}_j$). Ideally, negligible interference is caused to users outside both circles.}\label{figure_system_model}
\end{figure}

This coordination framework is characterized by the sets $\mathcal{C}_j,\mathcal{D}_j$, and the mnemonic rule is that $\mathcal{D}_j$ describes \emph{data} from transmitter $j$ while $\mathcal{C}_j$ describes \emph{coordination} from transmitter $j$. To reduce backhaul signaling of data, the cardinality of
$\mathcal{D}_j$ is typically smaller than that of $\mathcal{C}_j$. These
sets are illustrated in Fig.~\ref{figure_system_model} and can be selected based on long-term channel gains (see \cite{Bjornson2011a} for details).
Special cases include ideal network MIMO where all transmitters
serve all users (with
$\mathcal{C}_j\!=\!\mathcal{D}_j\!=\!\{1,\ldots,K_r\}$) and the interference channel (with $K_t\!=\!K_r$,
$\mathcal{D}_j\!=\!\{j\}$, and $\mathcal{C}_j\!=\!\{1,\ldots,K_r\}$). More detailed examples are available in \cite{Bjornson2012a} and \cite{Bjornson2011a}.

To enable coordinated transmissions, perfect phase coherence and synchronous interference is assumed between transmitters that serve users jointly (cf. \cite{Zhang2008a}).
The effective flat fading channel from $\textrm{BS}_j$ to
$\textrm{MS}_k$ is denoted $\vect{h}_{jk}$. The combined effective channel from all transmitters
is denoted $\vect{h}_{k}=[\vect{h}_{1k}^T \ldots \vect{h}_{K_t k}^T
]^T \in \mathbb{C}^{N \times 1}$ with $N=\sum_{j=1}^{K_t} N_j$. The received signal
at $\textrm{MS}_k$ is modeled as
\begin{equation} \label{eq_system_model}
y_{k}= \vect{h}_{k}^H \vect{C}_k \Big( \sum_{\bar{k}=1}^{K_r}
\vect{D}_{\bar{k}} \vect{s}_{\bar{k}} + \boldsymbol{\xi} \Big) + n_{k}
\end{equation}
where $\vect{D}_k$ selects the transmit antennas that send the zero-mean data signal $\vect{s}_{k} \in \mathbb{C}^{N \times 1}$ to $\textrm{MS}_k$.
The signal correlation matrix $\vect{S}_{k} = \mathbb{E}\{ \vect{s}_{k} \vect{s}_{k}^H \}$ is a parameter in the transmission design.

Let $\vect{I}_M,\vect{0}_M \in \mathbb{C}^{M \times
M}$ denote identity and zero matrices, respectively.
Then, $\vect{D}_k \in
\mathbb{C}^{N \times N}$ is block-diagonal and the $j$th block is
$\vect{I}_{N_j}$ if $k \in \mathcal{D}_j$ and $\vect{0}_{N_j}$ if $k
\not \in \mathcal{D}_j$.
Similarly, $\vect{C}_k \in \mathbb{C}^{N \times
N}$ selects signals from transmitters with non-negligible
channels to $\textrm{MS}_k$; $\vect{C}_k \in \mathbb{C}^{N \times N}$ is block-diagonal
and the $j$th block is $\vect{I}_{N_j}$ if $k \in \mathcal{C}_j$ and
$\vect{0}_{N_j}$ if $k \not \in \mathcal{C}_j$. The remaining (weak) interference and thermal noise is modeled by $n_k \in \mathcal{CN}(0,\sigma_k^2)$

In contrast to the prior works \cite{Gesbert2010a,Weingarten2006a,Bjornson2010c,Mochaourab2011a,Jorswieck2008b,Shang2010a,Mochaourab2012a,Mohseni2006a,Zhang2010a,Karipidis2010a,Wiesel2006a,Bjornson2012a,Bjornson2011a,Yu2007a,Annapureddy2011a,Zhang2008a},
we consider that physical transceivers suffer from hardware impairments (e.g., nonlinear amplifiers, phase noise, and IQ-imbalance) \cite{Studer2010a,Holma2011a,Zetterberg2011a}. The transmitter impairments are well-modeled by the Gaussian distortion term $\boldsymbol{\xi} \in \mathcal{CN}(\vect{0},\vect{\Xi})$ \cite{Studer2010a}.
The covariance matrix $\vect{\Xi} \in \mathbb{C}^{N \times N}$ is assumed diagonal and the distortion power at the $n$th antenna is proportional to the transmit power at this antenna: $\vect{\Xi} = \sum_{n=1}^{N} \sum_{k=1}^{K_r} \vect{T}_n
\vect{D}_{k} \vect{S}_{k} \vect{D}_{k}^H \vect{T}_n^H$, where the $n$th diagonal-element of $\vect{T}_n$ is $\kappa_n$ while all other elements are zero.
The proportionality constant $\kappa_n$ is known as the error vector magnitude (EVM) and typically satisfies $\kappa_n \in [0,0.15]$, depending on the quality of the transmitter hardware.

\subsection{Power Constraints}

The transmission is limited by $L$ linear power constraints
\begin{equation} \label{eq_power_constraints}
\sum_{k=1}^{K_r} \tr\{ \vect{Q}_{lk} \vect{S}_k \} \leq q_l,  \quad l=1,\ldots,L,
\end{equation}
where $\vect{Q}_{lk} \in \mathbb{C}^{N \times N}$ are Hermitian positive semi-definite matrices for all $l,k$.
To ensure that the total power is constrained and only is allocated to dimensions used for transmission, these matrices must satisfy two conditions: a)
$\vect{Q}_{lk} - \vect{D}_{k}^H \vect{Q}_{lk} \vect{D}_{k}$ is diagonal and b) $\sum_{l=1}^{L} \vect{Q}_{lk} \succ \vect{0}_N \, \forall k$.

A total power constraint ($L=1$) as well as per-base station ($L=K_t$) and
per-antenna constraints ($L=N$) can be expressed as \eqref{eq_power_constraints}; see examples in \cite{Bjornson2011a}.
The matrices $\vect{Q}_{lk}$ could be identical among users served by the same set of base stations, but it is also possible to have user-specific or area-specific soft-shaping constraints that limit the
interference generated in certain channel subspaces identified by $\vect{Q}_{lk}$ (e.g., to not disturb neighboring systems \cite{Huang2010a}).

\subsection{Performance Region}

Most common quality measures are monotonic functions of the
signal-to-interference-and-noise ratio (SINR); for example, achievable data
rate, mean square error (MSE), and bit/symbol error rate (BER/SER). For the system model in \eqref{eq_system_model}, the SINR at
$\textrm{MS}_k$ becomes\footnote{For notational simplicity, dirty-paper coding has not been included. It can however be used to presubtract interference at the transmitter-side, and the corresponding SINR expressions can be achieved using the same approach as in \cite{Yu2007a}.}
\begin{equation} \label{eq_SINR_DL_general}
\textrm{SINR}_{k} (\vect{S}_{1},\ldots,\vect{S}_{K_r}) =  \frac{
\vect{h}_{k}^H \vect{D}_k \vect{S}_{k} \vect{D}_k^H \vect{h}_{k}
}{\sigma_{k}^2 \!+\! \vect{h}_{k}^H \vect{C}_{k} ( \fracSum{\bar{k}
\neq k} \vect{D}_{\bar{k}} \vect{S}_{\bar{k}} \vect{D}_{\bar{k}}^H + \vect{\Xi})
\vect{C}_{k}^H \vect{h}_{k}}.
\end{equation}
Thus, optimizing the performance means selecting the transmit correlation matrices
$\vect{S}_k$ appropriately.

Herein, we model the performance of $\textrm{MS}_k$ by an arbitrary continuous and
strictly increasing monotonic performance function
$g_k(\textrm{SINR}_{k})$ that we seek to maximize\footnote{In case
of a strictly monotonic decreasing error measure
$\tilde{g}_k(\textrm{SINR}_{k})$ that should be minimized (e.g.,
MSE, BER, or SER), we can maximize $g_k(\textrm{SINR}_{k})=\tilde{g}_k(0)-\tilde{g}_k(\textrm{SINR}_{k})$ instead.}
for each user. For simplicity, we let $g_k(0)=0$.
We define the region of performance
outcomes for feasible transmit correlation matrices
$\vect{S}_1,\ldots,\vect{S}_{K_r}$:
\begin{definition}
The achievable \emph{performance region} $\mathcal{R} \subset
\mathbb{R}_+^{K_r}$ is
\begin{equation} \label{eq_performance_region}
\mathcal{R} = \big\{ \left(g_1(\textrm{SINR}_{1}), \ldots,
g_{K_r}(\textrm{SINR}_{K_r}) \right) : \,\,\, ( \vect{S}_1,\ldots,\vect{S}_{K_r} ) \in
\mathcal{S} \big\}
\end{equation}
where the set of feasible transmit strategies is
\begin{equation} \label{eq_feasible_transmit_strategies}
\begin{split}
\mathcal{S}=\Big\{ ( \vect{S}_1,\ldots,\vect{S}_{K_r}  ): \,\,\,
\vect{S}_k \succeq \vect{0}_N \, \, \forall k, \,\,\,\, \sum_{k=1}^{K_r} \tr\{
\vect{Q}_{lk} \vect{S}_k\} \leq q_l \,\, \forall l \Big\}.
\end{split}
\end{equation}
\end{definition}
This region describes the performance that can be simultaneously
achieved by the different users. It is a compact region since $\mathcal{S}$ is compact.
The interesting part of $\mathcal{R}$ is a subset of the outer boundary, called the Pareto boundary, where no user can improve its
performance without degrading others' performance:
\begin{definition} \label{def_boundaries}
The \emph{outer boundary} $\partial \mathcal{R}^{+} \subseteq
\mathcal{R}$ and the \emph{Pareto boundary} $\partial \mathcal{R} \subseteq
\mathcal{R}$ consist of all  $\vect{r} \in \mathcal{R}$ for which
there is no $\vect{r}' \in \mathcal{R} \!\setminus \! \{ \vect{r}
\}$ with $\vect{r}' > \vect{r}$ or $\vect{r}' \geq \vect{r}$, respectively (component-wise inequalities).
\end{definition}
All efficient outcomes of performance optimization lie on the Pareto
boundary. There are no simple expressions for $\partial \mathcal{R}$, but explicit and implicit characterization are derived in the following sections.

\section{Explicit Pareto Boundary Characterization} \label{section_explicit}

In this section, we derive a subset of the feasible transmit
strategies in $\mathcal{S}$ that can attain the complete Pareto boundary and
is explicitly characterized using $K_r+L-2$ parameters from $[0,1]$. Recalling that each strategy consists of
$K_r$ transmit correlation matrices of size $N \times N$, the derived necessary condition
constitutes a major reduction of the search space for Pareto
optimal strategies.

We will exploit the sufficiency of single-stream beamforming, proved similarly to \cite[Theorem 1]{Bjornson2011a}:
\begin{lemma} \label{lemma_rank_one_pareto_boundary}
Each point $\vect{r} \in \partial \mathcal{R}$ can be attained by
some transmit strategy $(\vect{S}^*_{1},\ldots,\vect{S}^*_{K_r}) \in
\mathcal{S}$ satisfying $\rank(\vect{S}^*_{k})\leq 1 \,\, \forall k$.
\end{lemma}

The lemma says that it is sufficient to consider transmit correlation matrices that are either rank-one or identically zero. Next, we propose such a strategy based on a set of parameters.

\begin{strategy} \label{strategy_explicit_precoding}
For some given non-negative parameters $\{\mu_{k}\}_{k=1}^{K_r}$ and $\{\lambda_l\}_{l=1}^{L}$, let the signal correlation matrix of user $k$ be $\vect{S}_{k} = p_k \vect{w}_k \vect{w}_k^H$ with
\begin{align} \label{eq_parameterization_beamformer_powerallocation}
\vect{w}_k &= \frac{\vect{\Psi}_k^{\dagger} \vect{D}_k^H \vect{h}_{k}}{\| \vect{\Psi}_k^{\dagger} \vect{D}_k^H \vect{h}_{k} \|} \\
\left[\begin{IEEEeqnarraybox*}[][ccc]{ccc} p_1 \, &
\ldots &
\,p_{K_r}%
\end{IEEEeqnarraybox*}\right] &=
\left[\begin{IEEEeqnarraybox*}[][ccc]{ccc} \gamma_1 \sigma_1^2 \, &
\ldots &
 \, \gamma_{K_r} \sigma_{K_r}^2%
\end{IEEEeqnarraybox*}\right] \vect{M}^{\dagger}
\end{align}
where
\begin{align}
\vect{\Psi}_k =& \Big(
\sum_{\bar{k}=1}^{K_r} \frac{\mu_{\bar{k}}}{\sigma_{\bar{k}}^2} \vect{D}_k^H
\vect{C}_{\bar{k}}^H \big(\vect{h}_{\bar{k}} \vect{h}_{\bar{k}}^H + \sum_{n=1}^N \vect{T}_n^H \vect{h}_{\bar{k}} \vect{h}_{\bar{k}}^H \vect{T}_n \big)
\vect{C}_{\bar{k}} \vect{D}_k \\ \notag
 &\quad \quad \quad +\sum_{l=1}^{L} \frac{\lambda_l}{q_l} \vect{Q}_{lk} \Big) \\
\gamma_k =& \frac{\mu_k}{\sigma_k^2} \vect{h}_{k}^H \vect{D}_k \big(\vect{\Psi}_k - \frac{\mu_{k}}{\sigma_k^2} \vect{D}_k^H
\vect{h}_{k} \vect{h}_{k}^H \vect{D}_k \big)^{\dagger} \vect{D}_k^H \vect{h}_{k}. \label{eq_gamma}
\end{align}
In \eqref{eq_parameterization_beamformer_powerallocation}-\eqref{eq_gamma}, we let $(\cdot)^{\dagger}$ denote the Moore-Penrose pseudoinverse and the $ij$th element
of $\vect{M} \in \mathbb{R}^{K_r \times K_r}$ is
\begin{equation} \label{eq_parameterization_M-matrix}
[ \vect{M} ]_{ij} = \begin{cases} | \vect{h}_{i}^H \vect{D}_i
\vect{w}_i|^2 - \gamma_i \fracSumtwo{n=1}{N} | \vect{h}_{i}^H \vect{D}_i \vect{T}_n \vect{w}_i|^2, & i = j, \\ - \gamma_j \big(| \vect{h}_{j}^H \vect{C}_{j}
\vect{D}_i \vect{w}_i|^2 + \fracSumtwo{n=1}{N} | \vect{h}_{j}^H \vect{C}_{j}
\vect{D}_i \vect{T}_n \vect{w}_i|^2 \big), & i \neq j. \end{cases}
\end{equation}
\end{strategy}

The following theorem shows that Strategy \ref{strategy_explicit_precoding} can attain any point on the Pareto boundary $\partial \mathcal{R}$ by proper parameter selection.

\begin{theorem} \label{theorem_explicit_parametrization}
Each Pareto optimal point $\vect{r} \in \partial \mathcal{R}$ is attained by $(\vect{S}_{1},\ldots,\vect{S}_{K_r}) \in \mathcal{S}$ given by Strategy \ref{strategy_explicit_precoding} for some selection of the parameters
$\{\mu_{k}\}_{k=1}^{K_r}$ and $\{\lambda_l\}_{l=1}^{L}$ that satisfies $\sum_{k=1}^{K_r} \mu_{k} =1$ and $\sum_{l=1}^{L} \lambda_l = 1$.
\end{theorem}
\begin{IEEEproof}
For brevity, the proof is given in the Appendix.
\end{IEEEproof}

The characterization in Theorem \ref{theorem_explicit_parametrization} uses $K_r+L$ parameters from $[0,1]$, but we only need to select $K_r+L-2$ parameters since the last
two are given by the two sum constraints. Observe that our novel characterization only has a single parameter per user, although multiple base stations are involved.
This reduces the number of parameters compared with the prior work in \cite{Bjornson2010c,Mochaourab2011a,Jorswieck2008b,Shang2010a} and \cite{Zhang2010a}, where each
transmitter has its own parameter for each user. The parameters $\mu_{k},\lambda_l$ implicitly determine beamforming directions and power
allocation. The direction $\vect{w}_k$ in \eqref{eq_parameterization_beamformer_powerallocation} is created by rotating
maximum ratio transmission $\vect{D}_k^H \vect{h}_k$ using the matrix $\vect{\Psi}_k$, whose terms determine to which extent power constraints, co-user interference, and hardware distortion are taken into account.
These terms are weighted by $\mu_{k},\lambda_l$ and their impact is showed next.

\begin{corollary} \label{cor_g_derivatives}
If $g_k(\cdot)$ is differentiable, changing the parameters in Theorem
\ref{theorem_explicit_parametrization} impacts the
performance of $\textrm{MS}_k$ as follows:
\begin{equation}
\begin{split}
 \frac{\partial}{\partial \mu_{\bar{k}}}g_k(\textrm{SINR}_k) &
\begin{cases} \geq 0, & k = \bar{k}, \\[-2mm]
 \leq 0, & k \neq \bar{k},
\end{cases} \\
 \frac{\partial}{\partial \lambda_{l}} g_k(\textrm{SINR}_k) & \leq 0
\quad \forall l.
\end{split}
\end{equation}
\end{corollary}
\begin{IEEEproof}
Observe that Strategy \ref{strategy_explicit_precoding} gives
$\textrm{SINR}_k = \gamma_k$. Differentiation of the $\gamma_k-$expression in Strategy \ref{strategy_explicit_precoding} proves the corollary, in conjunction with the monotonicity of $g_k(\cdot)$.
\end{IEEEproof}

The corollary proves that increasing $\mu_k$ improves the performance for
$\textrm{MS}_k$ and degrades it for other users, thus $\mu_k$ represents
the system priority of $\textrm{MS}_k$.
The level of enforcement of the $l$th power constraint is determined by $\lambda_l$ and should be small
to boost performance; $\lambda_l$ is zero for inactive constraints.
It is non-trivial which parameters to
modify to improve system performance and fulfill all constraints.
In fact, it is unlikely to find Pareto optimal points by trial-and-error selection.
However, close-to-optimal performance is relatively easy to achieve---the well-known
signal-to-leakage-and-noise ratio (SLNR)
beamforming \cite{Zhang2008a} and distributed virtual SINR (DVSINR) beamforming
\cite{Bjornson2010c} are achieved by simple parameter selections.

Strategy \ref{strategy_explicit_precoding} will not produce
feasible strategies for all parameter selections, but this is easily arranged.

\begin{strategy} \label{strategy_explicit_precoding_improved}
Let $(\vect{S}_{1},\ldots,\vect{S}_{K_r})$ be suggested by Strategy \ref{strategy_explicit_precoding}.
The modified strategy with
$\bar{\vect{S}}_k = \vect{S}_{k}/c \,\, \forall k$ and $c=\max_l (
\sum_k \tr\{ \vect{Q}_{lk} \vect{S}_k \}/ q_l )$ will always be feasible.
\end{strategy}

This modification only affects suboptimal and infeasible strategies since Pareto optimal strategies have $c=1$ (i.e., satisfy at least one constraint with equality,
see \cite[Theorem 2]{Bjornson2011a}).

\section{Implicit Pareto Boundary Characterization} \label{section_implicit}

\begin{figure}[t!]
\begin{center}
\includegraphics[width=\columnwidth]{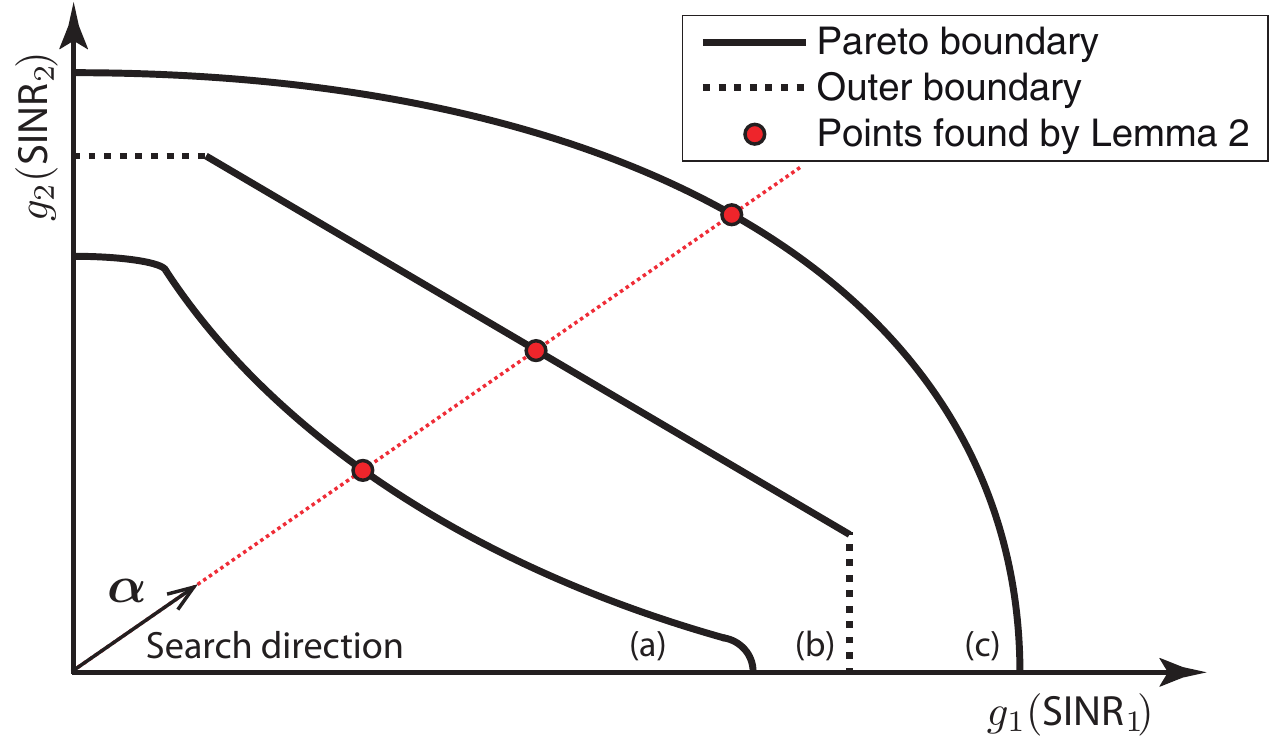}
\end{center}
\caption{Example of performance regions: (a) is non-convex and (b), (c) are convex. The outer boundary and Pareto boundary are identical, except in Case (b).
 The search direction $\boldsymbol{\alpha}$ of the implicit characterization is shown along with the Pareto optimal points that it will find.}\label{figure_boundary_examples}
\end{figure}

The explicit characterization in Section \ref{section_explicit} reduces the search-space for Pareto optimal points, but not all parameter selections attain the Pareto boundary. Next, we provide an implicit characterization that provides a point on the outer boundary for \emph{every} parameter selection, but at the expense of solving a
quasi-convex optimization problem. As pointed out in Definition \ref{def_boundaries}, the Pareto boundary is only a subset of the outer boundary. However, they are equal
in many multicell scenarios; three examples are given in Fig.~\ref{figure_boundary_examples} and only the flat parts (those orthogonal to an axis) in Case (b) are
not Pareto optimal.

The approach used herein is based on \emph{fairness-profiles} \cite{Bjornson2012a}, which is a conceptional extension of the rate-profile approach in \cite{Mohseni2006a} where users achieve pre-defined portions of the sum performance. We use $K_r-1$ parameters to define a fairness-profile $\boldsymbol{\alpha}=[\alpha_1,\ldots,\alpha_{K_r}] \in \mathbb{R}^{K_r}$ with non-negative entries and unit $L_1$-norm (thus, the last entry is given as $\alpha_{K_r}=1-\sum_{k=1}^{K_r-1} \alpha_k$). This vector points out the direction in which we search for the outer boundary; see Fig.~\ref{figure_boundary_examples}. The corresponding optimization problem is formulated as
\begin{equation} \label{eq_optproblem_performance-profile}
\begin{split}
\maximize{g_{\text{sum}}, \, (\vect{S}_{1},\ldots,\vect{S}_{K_r}) \in \mathcal{S}}\,\, & \quad g_{\text{sum}} \\
\mathrm{subject}\,\,\mathrm{to}\,\,\,\,\,\,\,\,\,\,\,
 & \quad g_k(\textrm{SINR}_k) = \alpha_k g_{\text{sum}} \quad
 \forall k.
\end{split}
\end{equation}
This problem might seem difficult, but since single-stream beamforming can be used (see Lemma
\ref{lemma_rank_one_pareto_boundary}) it can be formulated as quasi-convex.
By setting $\vect{h}_{k}^H \vect{D}_k \vect{v}_{k}>0$ (without loss of generality) and using bisection techniques \cite{Boyd2004a}, it can be solved with linear convergence as a series of convex feasibility problems:

\begin{lemma} \label{lemma_implicit_bisection}
The solution to \eqref{eq_optproblem_performance-profile} can be achieved by bisection over the range $\mathcal{G}=[0,g_{\text{max}}]$ of values for $g_{\text{sum}}$.
For a given $g_{\text{candidate}} \in \mathcal{G}$, the convex feasibility problem
\begin{align} \label{eq_expressed_as_socp}
\mathrm{find}\,\, & \,\, \vect{v}_1,\ldots,\vect{v}_{K_r} \\ \notag
\mathrm{subject}\,\,\mathrm{to}\,\, & \,\, \sum_{k=1}^{K_r} \vect{v}_k^H \vect{Q}_{lk} \vect{v}_k \leq q_l \quad
\forall l, \\ \notag
& \, \,\Im\{ \vect{h}_{k}^H \vect{D}_k \vect{v}_{k} \}=0 \quad \forall k,
\end{align}
\begin{equation} \notag
\begin{split}
&\sqrt{ \sigma_{k}^2 \!+\!   \fracSum{\bar{k}
\neq k} | \vect{h}_{k}^H \vect{C}_{k} \vect{D}_{\bar{k}}
\vect{v}_{\bar{k}} |^2 + \fracSumtwo{\bar{k}=1}{K_r} \fracSumtwo{n=1}{N} | \vect{h}_{k}^H \vect{C}_{k} \vect{D}_{\bar{k}}
\vect{T}_n \vect{v}_{\bar{k}} |^2 } \\ & \quad \quad \quad \quad \quad \quad \leq \frac{1}{\sqrt{\gamma_k}} \vect{h}_{k}^H \vect{D}_k \vect{v}_{k} \quad \forall k
\end{split}
\end{equation}
is solved for $\gamma_k=g_k^{-1}(\alpha_k g_{\textrm{candidate}})$. If there exist feasible solutions, all $g \in \mathcal{G}$ with $g< g_{\text{candidate}}$ are removed from $\mathcal{G}$. Otherwise, all $g \in \mathcal{G}$ with $g \geq g_{\text{candidate}}$ are removed from $\mathcal{G}$.

The initial upper bound can, for instance, be selected as
$g_{\text{max}} = \sum_{k=1}^{K_r} g_k ( \| \vect{D}_k^H \vect{h}_{k} \|^2 / (\nu_k \sigma_k^2) )$ where $1/\nu_k$ is an upper bound on the transmit power and is the smallest strictly positive eigenvalue of
$\frac{\vect{D}_k^H \vect{Q}_{lk} \vect{D}_k}{q_l \tr(\vect{D}_k)}$ among all $l$.
\end{lemma}

In a practical implementation, this algorithm halves the range $\mathcal{G}$ (by testing the feasibility at midpoints) until it is smaller than some
pre-defined accuracy. The feasibility problem in \eqref{eq_expressed_as_socp} is a second order cone program and thus solvable in polynomial time using software packages such as CVX \cite{cvx}. The implicit parametrization is established by the following theorem.

\begin{theorem} \label{theorem_implicit_parametrization}
Each point on the outer boundary (and Pareto boundary) is achieved by Lemma \ref{lemma_implicit_bisection} for a \emph{unique} $\boldsymbol{\alpha}$ with unit $L_1$-norm and non-negative entries. Each such $\boldsymbol{\alpha}$ gives a point on the outer boundary.
\end{theorem}
\begin{IEEEproof}
For brevity, the proof is given in the Appendix.
\end{IEEEproof}

A similar characterization was proposed in \cite{Zhang2010a} and \cite{Karipidis2010a} for MISO interference channels, thus we extend the
previous work to arbitrary multicell scenarios with arbitrary linear power constraints and hardware impairments. The fairness-profile optimization
in \eqref{eq_optproblem_performance-profile} is also a solution to $\mathrm{maximize} \min_k g_k(\textrm{SINR}_k) /\alpha_k$, which can be viewed as
optimizing the weighted worst-user performance. The special case of $\alpha_k=1/K_r \, \, \forall k$ corresponds to classic worst-user
optimization; see, for example, \cite{Wiesel2006a,Bjornson2012a}.

\section{Two Simple Multicell Examples} \label{section_multicell_examples}

Next, we exemplify the characterizations on two simple multicell scenarios that can
be described by our general framework.

\subsection{MISO Interference Channel}

The Pareto boundary when $\textrm{BS}_j$ only transmits to $\textrm{MS}_j$ (i.e., $K_t=K_r$)
has attracted much attention under per-base station power constraints \cite{Jorswieck2008b,Shang2010a,Mochaourab2012a,Zhang2010a}.
The state-of-the-art Pareto
characterizations require $K_t(K_r-1)$ parameters, while
Theorem \ref{theorem_explicit_parametrization} uses $2K_r-2$ parameters under these conditions.
Thus, our novel characterization is advantageous whenever $K_r\geq 3$, and the benefit increases rapidly with $K_r$.
In the special case of $K_r=2$, recent work in \cite{Mochaourab2012a} only requires a single parameter; a similar reduction is not possible in our general multicell characterization without removing the support for arbitrary power constraints.

\begin{figure}[t!]
\begin{center}
\includegraphics[width=\columnwidth]{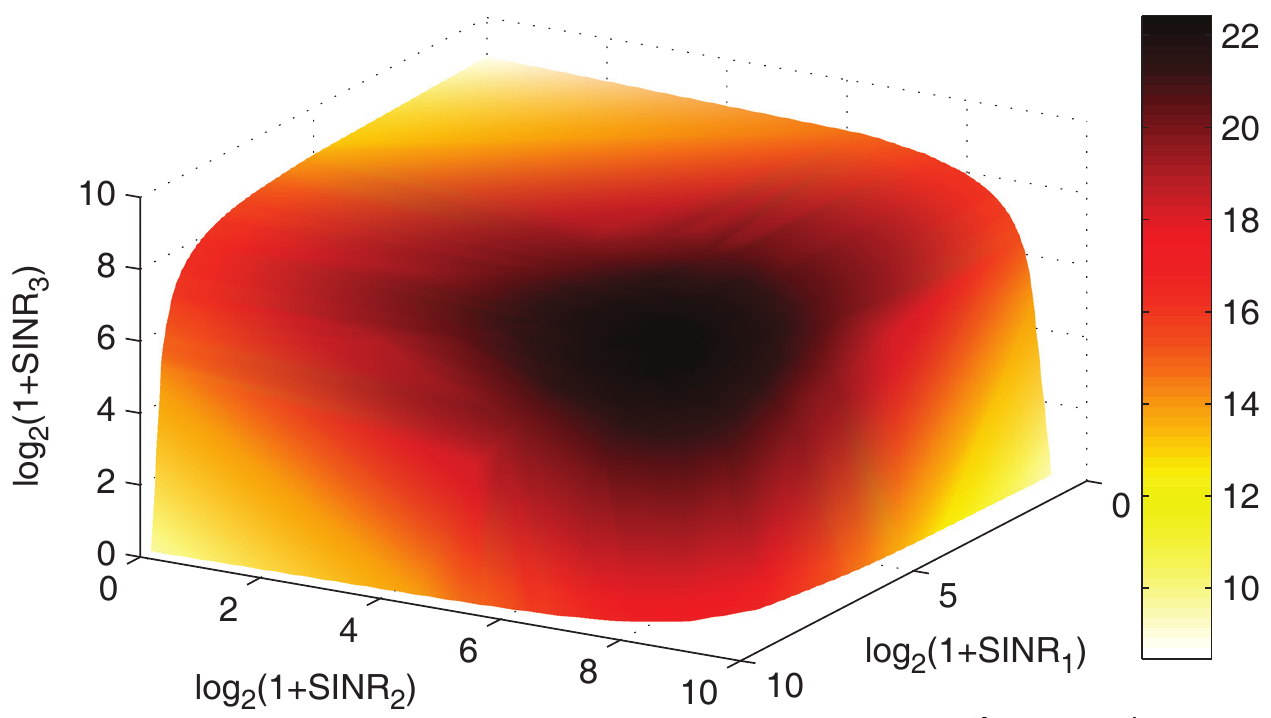}
\end{center}
\caption{Rate region (at SNR 10 dB) for a three-user MISO
interference channel with four antennas per transmitter, ideal hardware, and per-base
station constraints. The color bar shows the sum rate.}\label{figure_pareto3}
\end{figure}

In Fig.~\ref{figure_pareto3}, the performance region is shown for a uncorrelated Rayleigh fading channel realization with $K_t=K_r=3$, $N_t=4$, $\kappa_n=0$, and an average SNR of 10
dB (for maximum ratio transmission). The data rate $g_k(\textrm{SINR}_k) =
\log_2(1+\textrm{SINR}_k)$ is used as performance measure and the performance region is generated with Strategy \ref{strategy_explicit_precoding_improved}
by changing the four parameters in steps of $0.02$ and filling the space between the achieved points.
The region looks like a box with rounded edges, and the color bar shows the sum rate.

 \begin{figure}[t!]
\begin{center}
\includegraphics[width=\columnwidth]{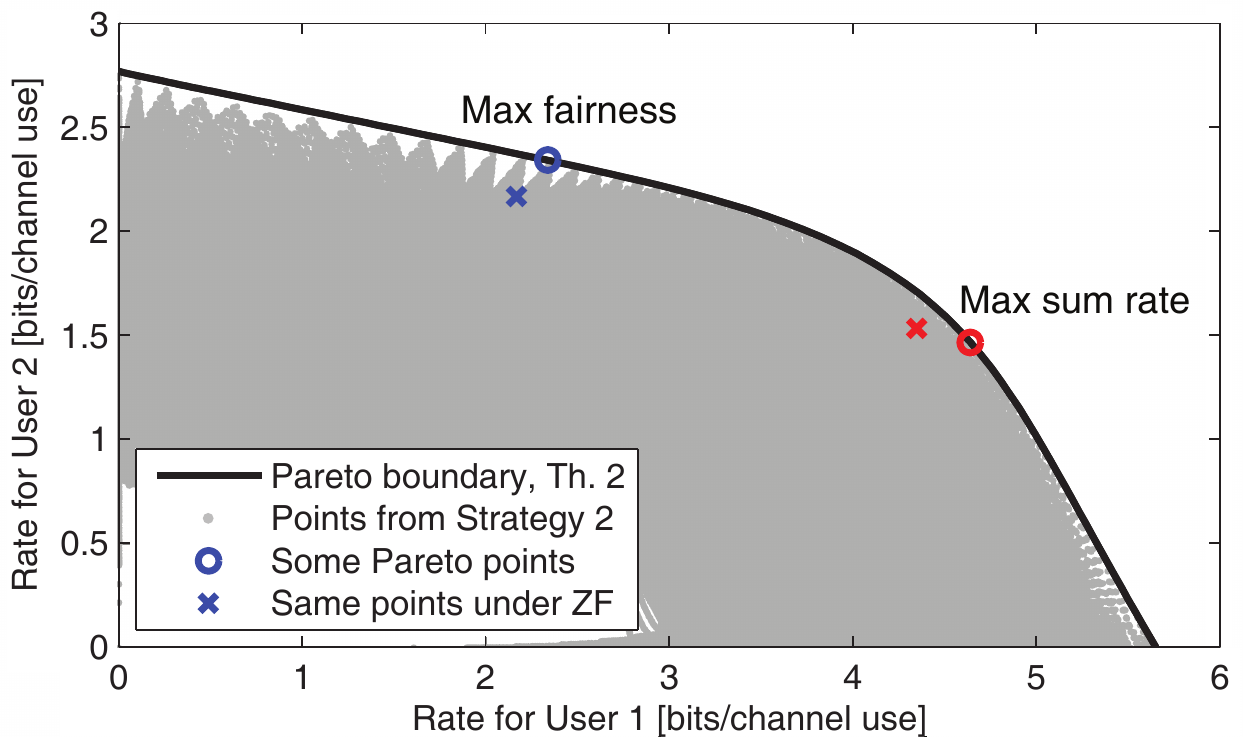}
\end{center}
\caption{Rate region (at SNR 10 dB) for ideal network MIMO
with two users, three transmit antennas, ideal hardware, and per-antenna constraints.}\label{figure_pareto1}
\end{figure}

\subsection{Ideal Network MIMO}

In this scenario, all base stations send data to all users and have
(identical) per-antenna power constraints as in \cite{Yu2007a}. The
performance region for this scenario has not been characterized in prior work.

\begin{figure}[t!]
\begin{center}
\includegraphics[width=\columnwidth]{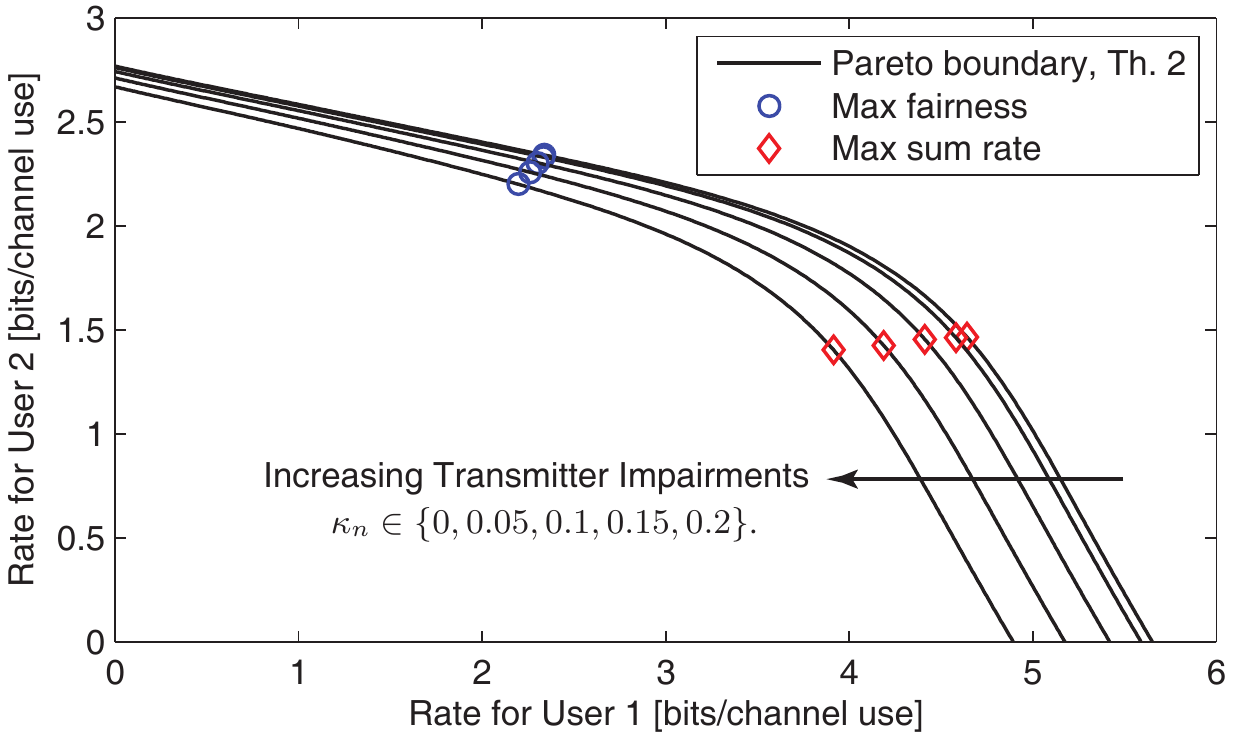}
\end{center}
\caption{Rate region (at SNR 10 dB) for the same scenario and channel realization as in Fig.~\ref{figure_pareto1}, but with varying transmitter
impairments: $\kappa_n \in \{0,0.05,0.1,0.15,0.2 \}$.}\label{figure_pareto1_impairments}
\end{figure}

We illustrate performance regions in a scenario with $N=3$ and
two single-antenna users. An uncorrelated Rayleigh fading channel realization is used
at an average SNR of 10 dB. The region with $g_k(\textrm{SINR}_k) =
\log_2(1+\textrm{SINR}_k)$ is shown in Fig.~\ref{figure_pareto1} for ideal hardware.
The shaded area is attained by Strategy \ref{strategy_explicit_precoding_improved} when
the three parameters are varied in steps of $0.01$. The outer
boundary of this area coincides with the plotted Pareto boundary, computed by Lemma \ref{lemma_implicit_bisection} using a grid of
$\boldsymbol{\alpha}$ vectors.
The maximal sum rate and fairness (i.e.,
$g_1=g_2$) points are shown, along with the corresponding points
under zero-forcing beamforming (based on \cite{Wiesel2008a}). The same
region is shown in Fig.~\ref{figure_pareto1_impairments} with impairments: $\kappa_n \in \{0,0.05,0.1,0.15,0.2 \}$. The sum rate loss is $0.2 \% - 13.5 \%$ depending the level of
impairments and which part of the boundary we consider; User 1 is more sensitive to impairments as this user has a stronger channel than User 2.

\begin{figure}[t!]
\begin{center}
\includegraphics[width=\columnwidth]{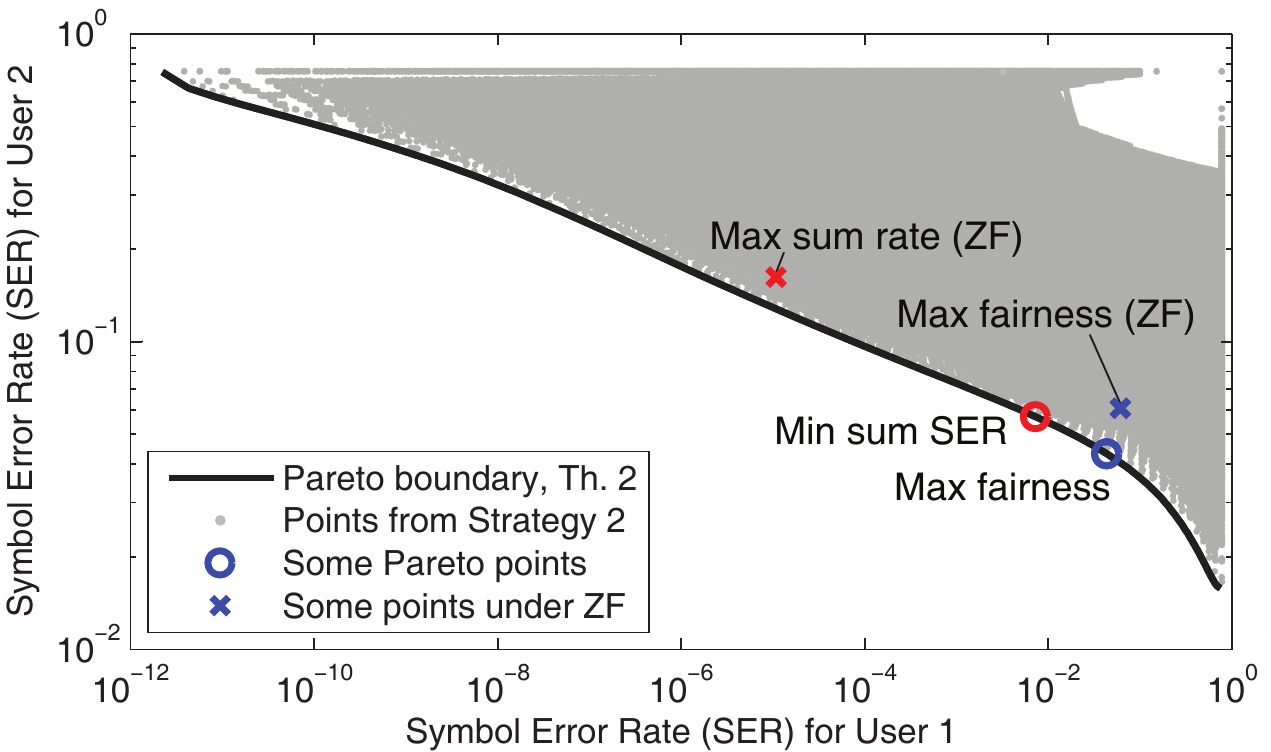}
\end{center}
\caption{Symbol error rate (SER) region (at SNR 10 dB) for ideal network
MIMO with two users, three transmit antennas, ideal hardware, and per-antenna
constraints.}\label{figure_pareto2}
\end{figure}

In Fig.~\ref{figure_pareto2}, the symbol error rates (SERs) with
4-QAM are shown for the same channel realization. The minimal sum SER and
maximal fairness are shown along with the same zero-forcing points
as in Fig.~\ref{figure_pareto1}. Comparing the two performance
measures, the maximal fairness points have the same interpretations in both cases while
optimizing sum performance yields very different results.

\section{Conclusion}

Efficient resource allocation is necessary to achieve the full potential of multicell multiantenna communication.
We considered a general setup where $K_t$ transmitters serve $K_r$
users under dynamic cooperation clusters, $L$ arbitrary power constraints, and practical hardware impairments.
The Pareto boundary of the performance region describes all efficient resource allocations.
A novel explicit characterization of the Pareto boundary was proposed that gives Pareto optimal transmit strategies in closed-form using $K_r+L-2$ parameters
between zero and one (instead of $K_t(K_r-1)$ as in prior work). We also extended previous work on implicit characterizations where each set of parameters guarantees
to find a point on the Pareto boundary, but at the expense of solving a quasi-convex optimization problem to find the corresponding transmit strategy.

\appendix

\subsubsection*{Proof of Theorem \ref{theorem_explicit_parametrization}}
\label{app_proof_theorem_explicit_parametrization}

For each given $\vect{r}=[r_1,\ldots,r_{K_r}] \in \partial
\mathcal{R}$, Lemma \ref{lemma_rank_one_pareto_boundary} states that
it is sufficient to consider rank-one solutions with $\vect{S}_{k} = \tilde{\vect{w}}_k \tilde{\vect{w}}_k^H$.
Observe that the optimal solution also solves the convex feasibility problem in
\eqref{eq_expressed_as_socp} when $\gamma_k=g_k^{-1}(r_k)$. Using
Lagrange multipliers  $\{\mu_{k}\}_{k=1}^{K_r}$ and $\{\lambda_l\}_{l=1}^{L}$, the Lagrangian of \eqref{eq_expressed_as_socp} can be expressed as (similarly to \cite[Proof of Proposition 1]{Yu2007a})
\begin{equation} \label{eq_lagrangian}
\mathcal{L} = \sum_{k=1}^{K_r} \mu_k - \sum_{l=1}^{L} \lambda_l + \sum_{k=1}^{K_r} \tilde{\vect{w}}_k^H \Big( \vect{\Psi}_k
- \frac{\mu_k}{\sigma_k^2} \big(1 + \frac{1}{\gamma_k} \big) \vect{D}_k^H
\vect{h}_{k} \vect{h}_{k}^H \vect{D}_k \Big) \tilde{\vect{w}}_k.
\end{equation}
The stationarity of the optimal
$\tilde{\vect{w}}_k$ (i.e., $\partial \mathcal{L}/\partial \tilde{\vect{w}}_k =
\vect{0}$) and multiplication with the Moore-Penrose pseudoinverse of $\vect{\Psi}_k$ gives
\begin{align} \label{eq_expression_for_wtilde_derivation}
\tilde{\vect{w}}_k =
\vect{\Psi}_k^{\dagger} \vect{D}_k^H
\vect{h}_{k} \underbrace{
\frac{\mu_k}{\sigma_k^2} \big(1 + \frac{1}{\gamma_k} \big)
\vect{h}_{k}^H
\vect{D}_k \tilde{\vect{w}}_k}_{=\textrm{scalar}}.
\end{align}
Since the phase of $\tilde{\vect{w}}_k$ will not affect the Lagrangian, $\tilde{\vect{w}}_k$ can without
loss of optimality be expressed as $\tilde{\vect{w}}_k=\sqrt{p_k}
\vect{w}_k$ with $\vect{w}_k$ as in
\eqref{eq_parameterization_beamformer_powerallocation} and for some $p_k \geq 0$.
To determine $p_k$ for $k=1,\ldots,K_r$, observe that since the
solution lies on the Pareto boundary, all SINR constraints in
\eqref{eq_expressed_as_socp} are satisfied with equality. The achieved SINRs $\gamma_k$ are
found by multiplying
\eqref{eq_expression_for_wtilde_derivation} with $\vect{h}_{k}^H \vect{D}_k$ from
the left and then divide by $\vect{h}_{k}^H \vect{D}_k \tilde{\vect{w}}_k$. The SINR equalities give $K_r$ linear equations that can be expressed and
solved as in \eqref{eq_parameterization_beamformer_powerallocation}.

Finally, observe that \eqref{eq_lagrangian} (and all equations in Strategy \ref{strategy_explicit_precoding}) is
unaffected by a common scaling of all Lagrange multipliers, thus we can assume $\sum_{k=1}^{K_r} \mu_{k} + \sum_{l=1}^{L} \lambda_l =2$
without losing any solutions. Also observe that $\sum_{k=1}^{K_r} \mu_k - \sum_{l=1}^{L} \lambda_l$ is the dual function and it is zero at the optimum due
to the strong duality of \eqref{eq_expressed_as_socp}. The combination of these two constraints gives that
$\sum_{k=1}^{K_r} \mu_{k} =1$ and $\sum_{l=1}^{L} \lambda_l =1$.

\subsubsection*{Proof of Theorem \ref{theorem_implicit_parametrization}}
\label{app_proof_theorem_implicit_parametrization}

Geometrically, \eqref{eq_optproblem_performance-profile} searches for the outer boundary along a ray from the origin in the direction of $\boldsymbol{\alpha}$.
To prove the first statement, observe that a ray can always be drawn to any outer boundary point, but the uniqueness of this ray requires a proof.
Say that the (outmost) intersection of a certain ray with the outer boundary is $\bar{\vect{r}}$, using the strategy $\vect{S}_k= \bar{p}_k \vect{w}_k \vect{w}_k^H \, \forall k$ with $\|\bar{\vect{w}}_k\|=1$.
Observe that every $\boldsymbol{\gamma}=[\gamma_1,\ldots,\gamma_{K_r}]^T \geq \vect{0}$ with $\boldsymbol{\gamma} < \bar{\vect{r}}$ can be
achieved by using the strategy $\vect{S}'_k=p_k \vect{w}_k \vect{w}_k^H \,\, \forall k$ with $p_k$ given by
\eqref{eq_parameterization_beamformer_powerallocation} (the existence of such $p_k$ and that $p_k\leq \bar{p}_k$ is easily
shown using interference functions \cite[Section 3]{Schubert2005a}).
Thus, all $\boldsymbol{\gamma}<\bar{\vect{r}}$ belong to the interior of $\mathcal{R}$ and therefore the ray can only intersect the outer boundary once.
The second part follows from the compactness of $\mathcal{R}$.

\section*{Acknowledgement}

The authors would like to thank P.~Zetterberg for sharing his knowledge on hardware impairments.

\bibliographystyle{IEEEtran}
\bibliography{IEEEabrv,refs}

\end{document}